# Breakdown of Three-dimensional Dirac Semimetal State in pressurized $Cd_3As_2$

Shan Zhang[1], Qi Wu[1], Leslie Schoop[2], Mazhar N. Ali[2], Youguo Shi[1], Ni Ni[3], Quinn Gibson[2], Shan Jiang[3], Vladimir Sidorov[4], Wei Yi[1], Jing Guo[1], Yazhou Zhou[1], Desheng Wu[1], Peiwen Gao[1], Dachun Gu[1], Chao Zhang[1], Sheng Jiang[5], Ke Yang[5], Aiguo Li[5], Yanchun Li[6], Xiaodong Li[6], Jing Liu[6], Xi Dai[1,7], Zhong Fang[1,7], Robert J. Cava[2]†, Liling Sun[1,7]†, Zhongxian Zhao[1,7]†

[1]Institute of Physics and Beijing National Laboratory for Condensed Matter Physics, Chinese Academy of Sciences, Beijing 100190, China

[2]Department of Chemistry, Princeton University, Princeton, NJ 08544, USA

[3]Department of Physics and Astronomy, UCLA, Los Angeles, CA90095, USA

[4]Institute for High Pressure Physics, Russian Academy of Sciences, 142190 Troitsk, Moscow, Russia

[5]Shanghai Synchrotron Radiation Facilities, Shanghai Institute of Applied Physics, Chinese Academy of Sciences, Shanghai 201204, China

[6] Institute of High Energy Physics, Chinese Academy of Sciences, Beijing 100049, China

[7]Collaborative Innovation Center of Quantum Matter, Beijing, 100190, China

We report the first observation of a pressure-induced breakdown of the 3D-DSM state in $Cd_3As_2$, evidenced by a series of *in-situ* high-pressure synchrotron X-ray diffraction (XRD) and single crystal transport measurements. We find that $Cd_3As_2$ undergoes a structural phase transition from a metallic tetragonal (T) phase in space group $I4_1/acd$ to a semiconducting monoclinic (M) phase in space group $P2_1/c$ at critical pressure 2.57 GPa, above this pressure, an activation energy gap appears, accompanied by distinct switches in Hall resistivity slope and electron mobility. These changes of crystal symmetry and corresponding transport properties manifest the breakdown of the 3D-DSM state in pressurized $Cd_3As_2$.

Three-dimensional Dirac semimetals (3D-DSMs) are new class of materials having non-trivial topology in their electronic states, featured by 3D Dirac points in the bulk and Fermi arcs on the surfaces.[1-4] These intriguing states predicted by theorists have been identified experimentally in real materials. $Cd_3As_2$ and $Na_3Bi$ are the materials with such a unique electronic structure.[5-11] Their conduction and valence bands contact at 3D Dirac points in momentum space where Dirac fermions disperse linearly along all momentum directions, resulting in the band gap closing in the bulk. Consistent with in this picture, $Cd_3As_2$ has high mobilities for electrons in bulk crystals,[12-19] so that it is expected to be a promising candidate in device applications.[20,21]

Similar to topological insulators, the 3D-DSM state is protected by either time reversal symmetry (TRS) or crystal rotational symmetry (CRS) along the z-axis.[1,2,22-27] On the breaking of TRS or CRS, the 3D-DSM state may be tuned into a variety of new quantum states with novel physical properties.[28-33] The tuning effect may be realized by either applying pressure or chemical doping. Of these two ways, pressure is an ideal method for finely tuning a system from one state to another without introducing chemical complexity; therefore the behavior of 3D-DSMs under pressure is of current research interest. In this study, we demonstrate that the pressure-induced alternation of CRS leads to the breakdown of the 3D-DSM state in single crystal $Cd_3As_2$ and how the transport properties change correspondingly.

High quality single crystals of $Cd_3As_2$ were grown by the flux method, as described in Ref. 34. Pressure was generated by a diamond anvil cell (DAC) with two opposing anvils sitting on a Be-Cu supporting plate. Diamond anvils with 500 μm and 300 μm flats were used for this study. A non-magnetic rhenium gasket with 200 μm and 100 μm diameter holes was used for different runs of the high-pressure studies. The four-probe method was applied in the (112) cleavage plane of a single crystal of $Cd_3As_2$ for all high pressure transport measurements. To keep the sample in a quasi-hydrostatic pressure environment, NaCl powder was employed as the pressure medium for the high-pressure resistance and Hall measurements. The high-pressure heat capacity ($C_{ac}$) of $Cd_3As_2$ was derived from *ac* calorimetry.[35,36] High pressure X-ray diffraction (XRD) experiments were performed at beam line 15U at the Shanghai Synchrotron Radiation Facility and at beam line 4W2 at the Beijing Synchrotron Radiation Facility, respectively. Diamonds with low birefringence were selected for the experiments. A monochromatic X-ray beam with a wavelength of 0.6199 Å was adopted for all measurements. Pressure was determined by the ruby fluorescence method.[37]

We first performed high-pressure powder X-ray diffraction measurements to detect the structural stability of $Cd_3As_2$ under pressure, because the 3D-DSM state is protected by the CRS along the z-axis.[1,2,27,31] As shown in Fig.1a, $Cd_3As_2$ has a tetragonal unit cell in the tetragonal I4$_1$/*acd* space group at ambient pressure, as reported in Ref. 34. Applying pressure up to 2.42 GPa, no new peaks appear over the range of 2θ angles studied. Instead, pressure shifts each Bragg peak to a higher angle, reflecting a continuous reduction of the *d* spacing in the tetragonal (T) phase. By pressures of ~3.78 GPa and above, a set of new diffraction peaks is clearly visible, demonstrating that the ambient-pressure crystal symmetry is destroyed and a different crystal structure is present at high pressures. High-pressure resistance measurements indicate that the precise critical pressure for the phase transition is 2.57 GPa (Fig.1b). To trace the stability of the two phases (T phase and new high pressure phase) as temperature changes, we performed high-pressure specific heat measurements at 2.3 GPa and 4.1 GPa, respectively. We find that, once formed, both the T phase and the new high pressure phase are stable under the temperatures ranging from 300 K to 4 K (Fig.1c).

Years ago, the pressure-induced structural phase transition in $Cd_3As_2$ was studied by X-ray diffraction camera (XRDC)[38] and differential thermal analysis (DTA),[39] respectively. The XRDC measurements showed that $Cd_3As_2$ transforms into a trigonal structure in space group $P\bar{3}m1$ at pressures near 2.5 GPa,[38] while the DTA results proposed that $Cd_3As_2$ undergoes a crystal phase transition at ~ 1.7 GPa but no XRD data is provided.[39] However, as shown in Fig.1d, most of our X-ray diffraction peaks of the pressurized $Cd_3As_2$ phase (the peak positions are indicated by short green lines) do not come close to the peak positions calculated from the proposed trigonal symmetry cell (as indicated by short black lines).[38] We thus do the refinements for the high-pressure X-ray diffraction data and find that the high pressure phase is monoclinic (M) in space group $P2_1/c$ (Fig.1d and 1e). The pressure dependence of lattice parameters in the T and M phases can be found in Fig.S1 (Supplementary Information). The obtained results indicate that pressure induces a change of CRS from a space group based on 4-fold rotational symmetry to the one based on 2-fold rotational symmetry.

High pressure transport measurements reveal details of the property changes before and after the crystal structure phase transition. Figure 2 shows the temperature dependent resistance at different pressures. As can be seen, pressure destabilizes the metallic behavior of $Cd_3As_2$, converting the sample from its ambient pressure metallic behavior to semiconducting behavior between 2.45 and 3.99 GPa (Fig.2a). The change from metallic to semiconducting state lies in the pressure range of the T-M phase transition. On further increasing pressure, the semiconducting behavior becomes pronounced (Fig.2b). We estimate the activation energy gap ($\varepsilon_A$) for the excitation of charge carriers as a function of pressure, on the basis of the Arrhenius equation $\rho \sim exp(\varepsilon_A/2k_BT)$ (inset of Fig.2b). The $\varepsilon_A$ opens at 3.99 GPa ($\varepsilon_A$ =13.5 meV) in the M phase and increases with pressure rapidly. This phenomenon is in stark contrary to the common picture that pressure usually drives an insulator (or semiconductor) into a metal due to the band broadening, which leads us to expect that the semimetal $Cd_3As_2$, after undergoing a structure phase transition under pressure, will be more metallic. This unusual pressure effect on $Cd_3As_2$ demonstrates the uncommon characteristic of its electron structure, which should be stemmed from its unique topological semimetal state protected by CRS. The gap opening found in this study gives strong support for that the pressure-induced change of CRS splits the bulk 3D Dirac points at the Fermi level, namely the breakdown of the 3D-DSM state.

We measured the Hall resistance ($R_{XY}$) as a function of magnetic field at 4 K under pressure. Figure 3a and 3b show the overall features of the $R_{XY}$ measured on the

$Cd_3As_2$ single crystal in the field to 4 T for various pressures. We find that $R_{XY}$ against magnetic field displays linear behavior both in its T phase[12-14] and M phase. However, the $R_{XY}$-$H$ curve takes on two distinct slopes: a large slope for the T phase and a small slope for the M phase (Fig.3c). Around the phase transition, the $dR_{XY}/dH$ drops by about 60%. The switch in Hall resistance slope supports the conclusion that the 3D-DSM state of $Cd_3As_2$ is destroyed with pressure because the change of $dR_{XY}/dH$ corresponds to a change of band structure.[40] The pressure dependence of $R_{XY}$ is plotted in Fig.3d, which clearly illustrates the remarkable difference of Hall resistance before and after the transition. The $R_{XY}$ of the pressurized $Cd_3As_2$ drops by more than a factor of 3 when it turns into the high pressure semiconducting phase. These data are the supporting evidence for breakdown of the 3D-DSM state in pressurized $Cd_3As_2$. Intriguingly, the $R_{XY}$ measured at 2.09 GPa shows kinks in both the positive and the negative directions of the magnetic field (Fig.3a), a possible indication of the presence of an intermediate electronic state that may be of significant interest for further study.

It has been shown that one of the striking features of the 3D-DSM is that the electron mobility is very high by virtue of 3D massless Dirac or Weyl fermions.[12-18] Experimentally, the best way to identify the breakdown of the 3D-DSM state under pressure is to measure the pressure dependence of the electron mobility. A remarkable drop in mobility should be observed when the 3D-DSM state is destroyed. To further clarify this issue, we obtained the pressure dependent electron mobility of $Ca_3As_2$ (Fig.4). We find that the mobility ($\mu$) of the sample in the tetragonal 3D-DSM state has a relatively high value, in good agreement with the values reported in the literature,[12,13,19,41] and is nearly pressure independent, revealing that the charge carriers in the pressure range from ambient pressure to 2.5 GPa are relatively massless. However, a dramatic reduction in electron mobility is found at pressures greater than 2.5 GPa; the mobility $\mu$ drops by about 74% at 3 GPa and by about 99% at 4 GPa. These results presented in Fig.4 are the further evidence of the breakdown of the 3D-DSM state in pressurized $Cd_3As_2$.

In summary, we find experimental evidence for a pressure-induced breakdown of the 3D-DSM state in $Cd_3As_2$ through *in-situ* high pressure X-ray diffraction and transport measurements. Our results demonstrate that external pressure leads $Cd_3As_2$ to undergo a crystal structure phase transition from a tetragonal phase in space group $I4_1/acd$ to a monoclinic phase in space group $P2_1/c$, which turns it from a Dirac semimetal into a semiconductor. This was characterized by the opening of an activation energy gap and substantial changes in the $dR_{XY}/dH$ slope and electron mobility. Further study on the breakdown of the 3D DSM state in $Cd_3As_2$ under pressure will be of significant interest.

**References**

1. Z. J. Wang, Y. Sun, X. Q. Chen, C. Franchini, G. Xu, H. M. Weng, X. Dai, and Z. Fang, Phys. Rev. B **85,** 195320 (2012).

2. Z. J. Wang, H. M. Weng, Q. S. Wu, X. Dai, and Z. Fang, Phys. Rev. B **88,** 125427 (2013).

3. M. Z. Hasan, and J. E. Moore, Annu. Rev. Condens. Matter Phys. **2,** 55-78 (2011).

4.  S. M. Young, S. Zaheer, J. C. Y. Teo, C. L. Kane, E. J. Mele, and A. M. Rappe, Phys. Rev. Lett. **108,** 140405 (2012).

5.  Z. K. Liu, B. Zhou, Z. J. Wang, H. M. Weng, D. Prabhakaran, S. -K. Mo, Y. Zhang, Z. X. Shen, Z. Fang, X. Dai, Z. Hussain, and Y. L. Chen, Science **343,** 864-867 (2014).

6.  S. -Y. Xu, C. Liu, S. K. Kushwaha, T. -R. Chang, J. W. Krizan, R. Sankar, C. M. Polley, J. Adell, T. Balasubramanian, K. Miyamoto, N. Alidoust, G. Bian, M. Neupane, I. Belopolski, H. -T. Jeng, C. -Y. Huang, W. -F. Tsai, H. Lin, F. C. Chou, T. Okuda, A. Bansil, R. J. Cava, and M. Z. Hasan, arXiv:1312.7624.

7.  Z. K. Liu, J. Jiang, B. Zhou, Z. J. Wang, Y. Zhang, H. M. Weng, D. Prabhakaran, S-K. Mo, H. Peng, P. Dudin, T. Kim, M. Hoesch, Z. Fang, X. Dai, Z. X. Shen, D. L. Feng, Z. Hussain, and Y. L. Chen, Nat. Mater. **13,** 677–681 (2014).

8.  M. Neupane, S. –Y. Xu, R. Sankar, N. Alidoust, G. Bian, C. Liu, I. Belopolski, T. –R. Chang, H. –T. Jeng, H. Lin, A. Bansil, F. C. Chou, and M. Z. Hasan, Nat. Commun. **5,** 3786 (2014).

9.  S. Borisenko, Q. Gibson, D. Evtushinsky, V. Zabolotnyy, B. Büchner, and R. J. Cava, Phys. Rev. Lett. **113,** 027603 (2014).

10. H. M. Yi, Z. J. Wang, C. Y. Chen, Y. G. Shi, Y. Feng, A. J. Liang, Z. J. Xie, S. L. He, J. F. He, Y. Y. Peng, X. Liu, Y. Liu, L. Zhao, G. D. Liu, X. L. Dong, J. Zhang, M. Nakatake, M. Arita, K. Shimada, H. Namatame, M. Taniguchi, Z. Y. Xu, C. T. Chen, X. Dai, Z. Fang, and X. J. Zhou. Sci. Rep. **4,** 6106 (2014).

11. S. Jeon, B. B. Zhou, A. Gyenis, B. E. Feldman, I. Kimchi, A. C. Potter, Q. D. Gibson, R. J. Cava, A. Vishwanath, and A. Yazdani, Nat. Mater. **13,** 851-856 (2014).

12. T. Liang, Q. Gibson, M. N. Ali, M. H. Liu, R. J. Cava, and N. P. Ong, Nat. Mater. doi:10.1038/nmat4143.

13. L. P. He, X. C. Hong, J. K. Dong, J. Pan, Z. Zhang, J. Zhang, and S. Y. Li, Phys. Rev. Lett. **113,** 246402 (2014).

14. J. Y. Feng, Y. Pang, D. S. Wu, Z. J. Wang, H. M. Weng, J. Q. Li, X. Dai, Z. Fang, Y. G. Shi, and L. Lu, arXiv:1405.6611.

15. Y. F. Zhao, H. W. Liu, C. L. Zhang, H. C. Wang, J. F. Wang, Z. Q. Lin, Y. Xing, H. Lu, J. Liu, Y. Wang, S. J, X. C. Xie, and J. Wang, arXiv:1412.0330.

16. A. Narayanan, M. D. Watson, S. F. Blake, Y. L. Chen, D. Prabhakaran, B. Yan, N. Bruyant, L. Drigo, I. I. Mazin, C. Felser, T. Kong, P. C. Canfield, and A. I. Coldea, arXiv:1412.4105.

17. J. Z. Cao, S. H. Liang, C. Zhang, Y. W. Liu, J. W. Huang, Z. Jin, Z. G. Chen, Z. J. Wang, Q. S. Wang, J. Zhao, S. Y. Li, X. Dai, J. Zou, Z. C. Xia, L. Li, and F. X. Xiu, arXiv:1412.0824.

18. S. E. R. Hiscocks, and C. T. Elliott, J. Mater. Sci. **4,** 784-788 (1969).

19. L. G. Caron, J.-P. Jay-Gerin, and M. J. Aubin, Phys. Rev. B **15,** 3879-3887 (1977).

20. K. C. Saraswat, C. O. Chui, D. Kim, T. Krishnamohan, and A. Pethe, Electron Devices Meeting, IEDM' 06, International (2006).

21. C. C. Stoumpos, C. D. Malliakas, and M. G. Kanatzidis, Inorg. Chem. **52,** 9019–9038 (2013).

22. L. Fu, C. L. Kane, and E. J. Mele, Topological Insulators in Three Dimensions, Phys. Rev. Lett. **98,** 106803 (2007).

23. D. Hsieh, D. Qian, L. Wray, Y. Xia, Y. S. Hor, R. J. Cava, and M. Z. Hasan, Nature **452,** 970-974 (2008).

24. M. Z. Hasan, and C. L. Kane, Rev. Mod. Phys. **82,** 3045 (2010).

25. X. L. Qi, and S. C. Zhang, Rev. Mod. Phys. **83,** 1057 (2011).

26. M. Dzero, K. Sun, V. Galitski, and P. Coleman, Phys. Rev. Lett. **104,** 106408 (2010).

27. B. J. Yang and N. Nagaosa, Nat. Commun. **5**, 4898 (2014).

28. A. A. Burkov, M. D. Hook, and L. Balents, Phys. Rev. B **84,** 235126 (2011).

29. G. B. Halász, and L. Balents, Phys. Rev. B **85,** 035103 (2012).

30. L. A. Wray, S. −Y. Xu, Y. Q. Xia, D. Hsieh, A. V. Fedorov, Y. S. Hor, R. J. Cava, A. Bansil, H. Lin, and M. Z. Hasan, Nat. Phys. **7,** 32-37 (2011).

31. W. Witczak-Krempa, G. Chen, Y. B. Kim, and L. Balents, Annu. Rev. Condens. Matter Phys. **5,** 57-82 (2014).

32. X. G. Wan, A. M. Turner, A. Vishwanath, and S. Y. Savrasov, Phys. Rev. B **83,** 205101 (2011).

33. B. −J. Yang, and N. Nagaosa, Nat. Commun. **5,** 4898 (2014).

34. M. N. Ali, Q. Gibson, S. Jeon, B. B. Zhou, A. Yazdani, and R. J. Cava, Inorg. Chem. **53,** 4062-4067 (2014).

35. A. Eichler, and W. Gey, Rev. Sci. Instrum. **50,** 1445-1452 (1979).

36. V. A. Sidorov, E. D. Bauer, N. A. Frederick, J. R. Jeffries, S. Nakatsuji, N. O. Moreno, J. D. Thompson, M. B. Maple, and Z. Fisk, Phys. Rev. B **67**, 224419 (2003).

37. H. K. Mao, J. Xu, and P. M. Bell, J. Geophys. Res. **91,** 4673-4676 (1986).

38. M. D. Banus, and M. C. Lavine, High Temp. - High Press. **1,** 269-276 (1969).

39. C. W. F. T. Pistorius, High Temp. - High Press. **7,** 441-449 (1975).

40. X. H. Zhang, L. Q. Yu, S. V. Molnár, Z. Fisk, and P. Xiong, Phys. Rev. Lett. **103,** 106602 (2009).

41. J. Cisowski, and W. Zdanowicz, Acta. Phys. Pol. A **A43,** 295-299 (1977).

42. A. de. Combarieu, and J. -P. Jay-Gerin, Phys. Rev. B **25,** 2923 (1982).

**Acknowledgements**

The work in China was supported by the NSF of China (Grant No. 91321207, 11427805), 973 projects (Grant No.2011CBA00100 and 2010CB923000) and the Strategic Priority Research Program (B) of the Chinese Academy of Sciences (Grant No. XDB07020300). The work in the USA was supported by the ARO MURI on topological insulators, grant W911NF-12-1-0461 and the U.S. Department of Energy, Office of Science, Office of Basic Energy Sciences under Award Number DE-SC0011978.

†To whom correspondence should be addressed.

E-mail: llsun@iphy.ac.cn, rcava@Princeton.EDU and zhxzhao@iphy.ac.cn.

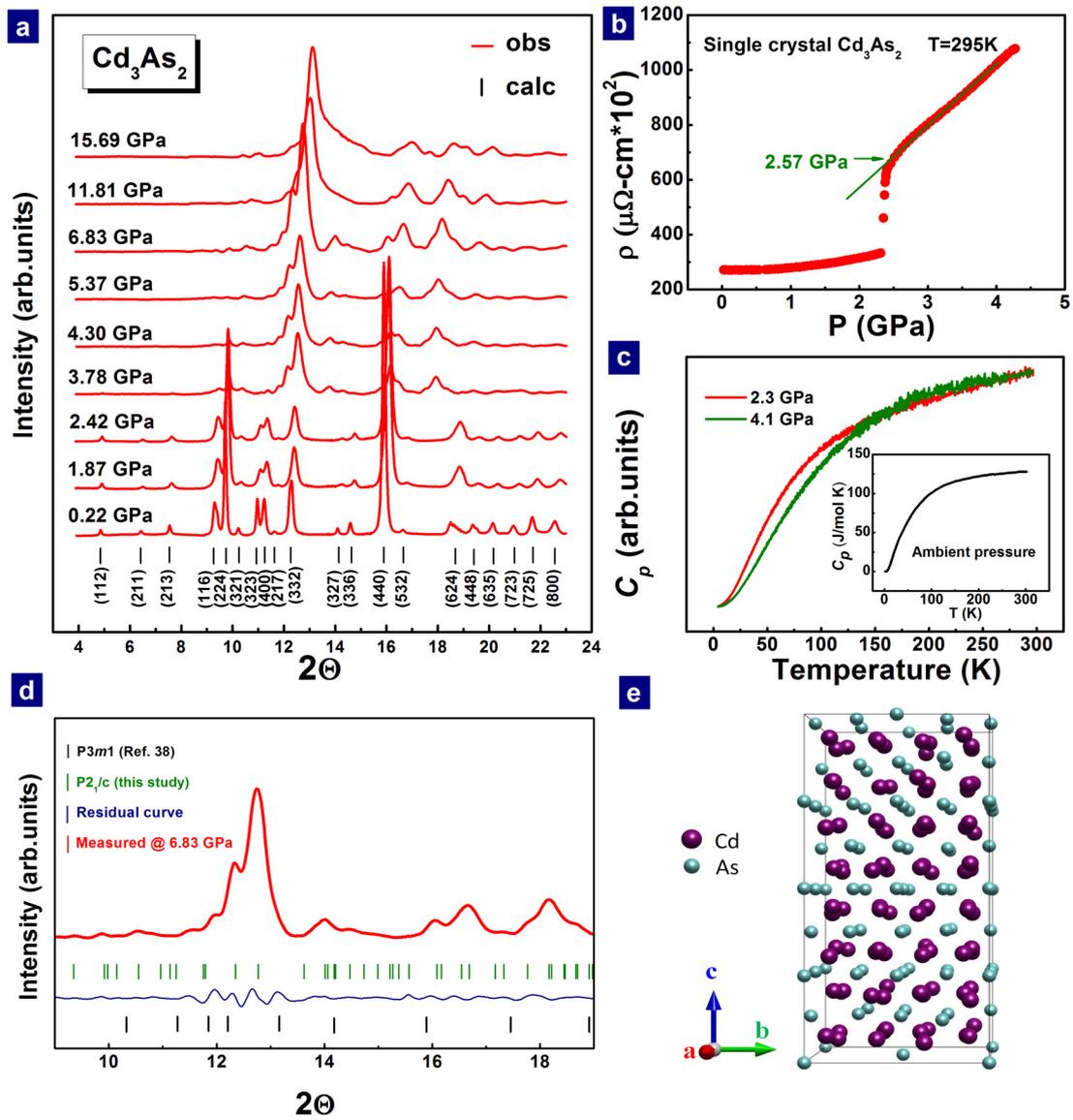

Figure 1 (a) X-ray powder diffraction patterns of $Cd_3As_2$ at various pressures at room temperature. The data below 2.42 GPa can be indexed well as the tetragonal (T) phase in space group $I4_1/acd$ (as indicated by black short lines), while on increasing pressure to 3.78 GPa and above, the data shows a set of new peaks, indicating a phase transition from the T phase to the high pressure phase. (b) Pressure dependence of electrical resistance for single crystal $Cd_3As_2$. The onset pressure of the transition is located at 2.57 GPa. (c) Temperature dependence of the heat capacity of a single crystal of $Cd_3As_2$ at 2.3 and 4.1 GPa. The heat capacity measured at different pressures decreases smoothly with decreasing temperature, demonstrating that no structural phase transitions occur in the temperature range of 4K-295K. The inset shows the heat capacity as a function of temperature measured at ambient pressure, which is in good agreement with the results of Ref.42. (d) Powder XRD data for $Cd_3As_2$ measured at 6.38 GPa, showing that the high pressure phase is the monoclinic one in space group $P2_1/c$. (e) Three dimensional structure of monoclinic phase.

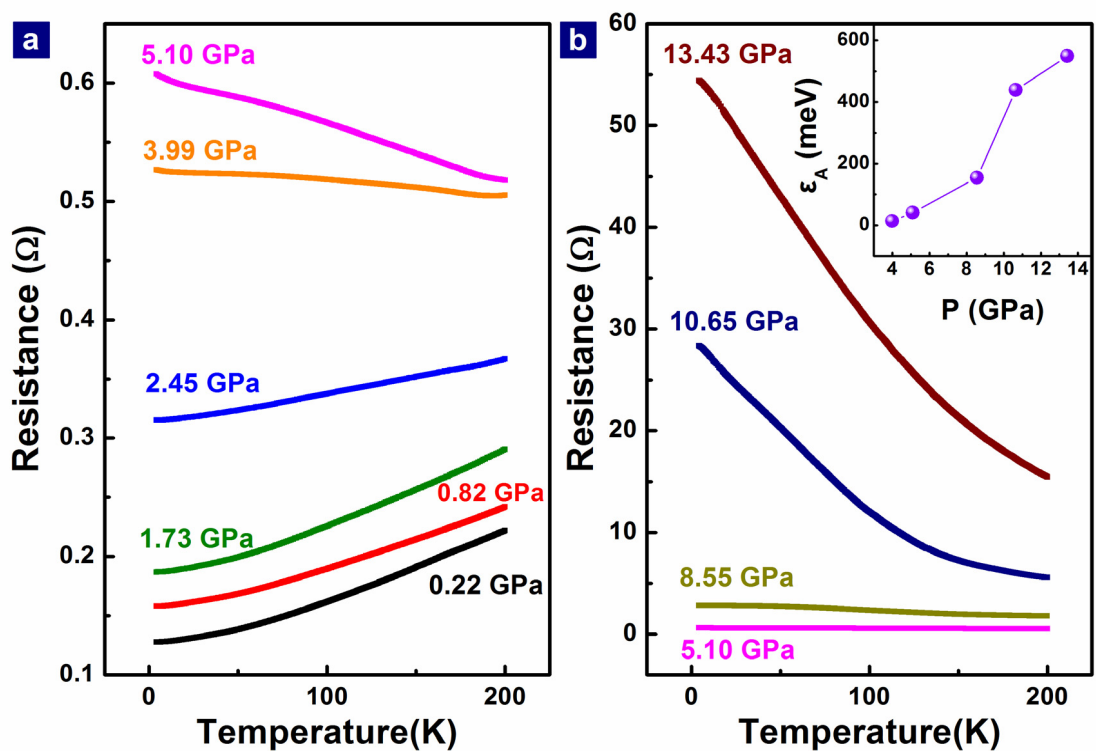

Figure 2 Temperature dependence of electrical resistance for single crystal $Cd_3As_2$ at different pressures. The inset of Fig.2 (b) displays the activation energy gap $\varepsilon_A$ as a function of pressure.

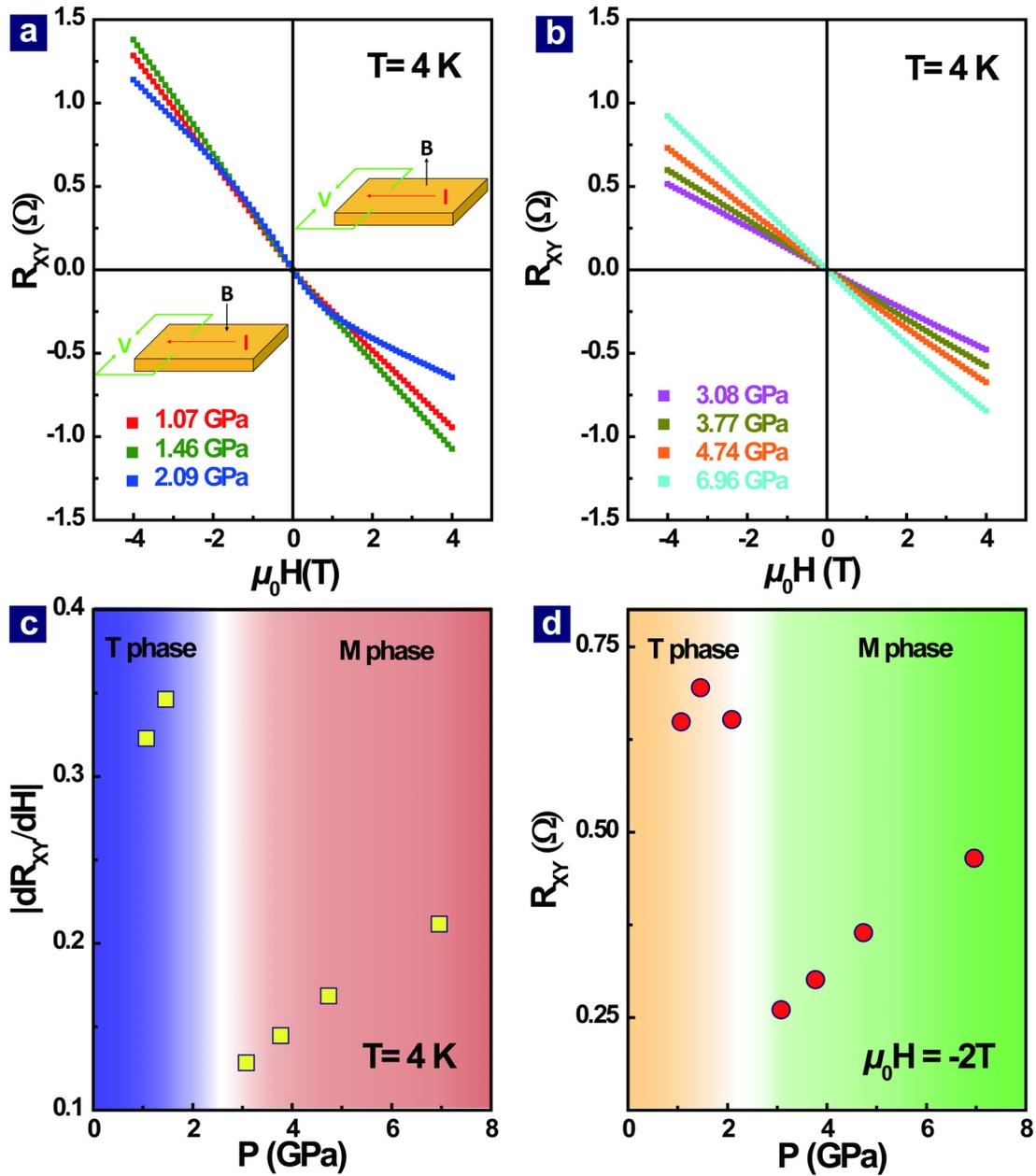

Figure 3 (a) and (b) Hall resistance ($R_{XY}$) as a function of magnetic field for single crystal $Cd_3As_2$ at different pressures. (c) and (d) $dR_{XY}/dH$ and $R_{XY}$ as a function of pressure, showing the dramatic drop in $dR_{XY}/dH$ and $R_{XY}$ (at a field of 2 Tesla) at 4 K near the transition from the Dirac semimetal to high pressure semiconducting phase.

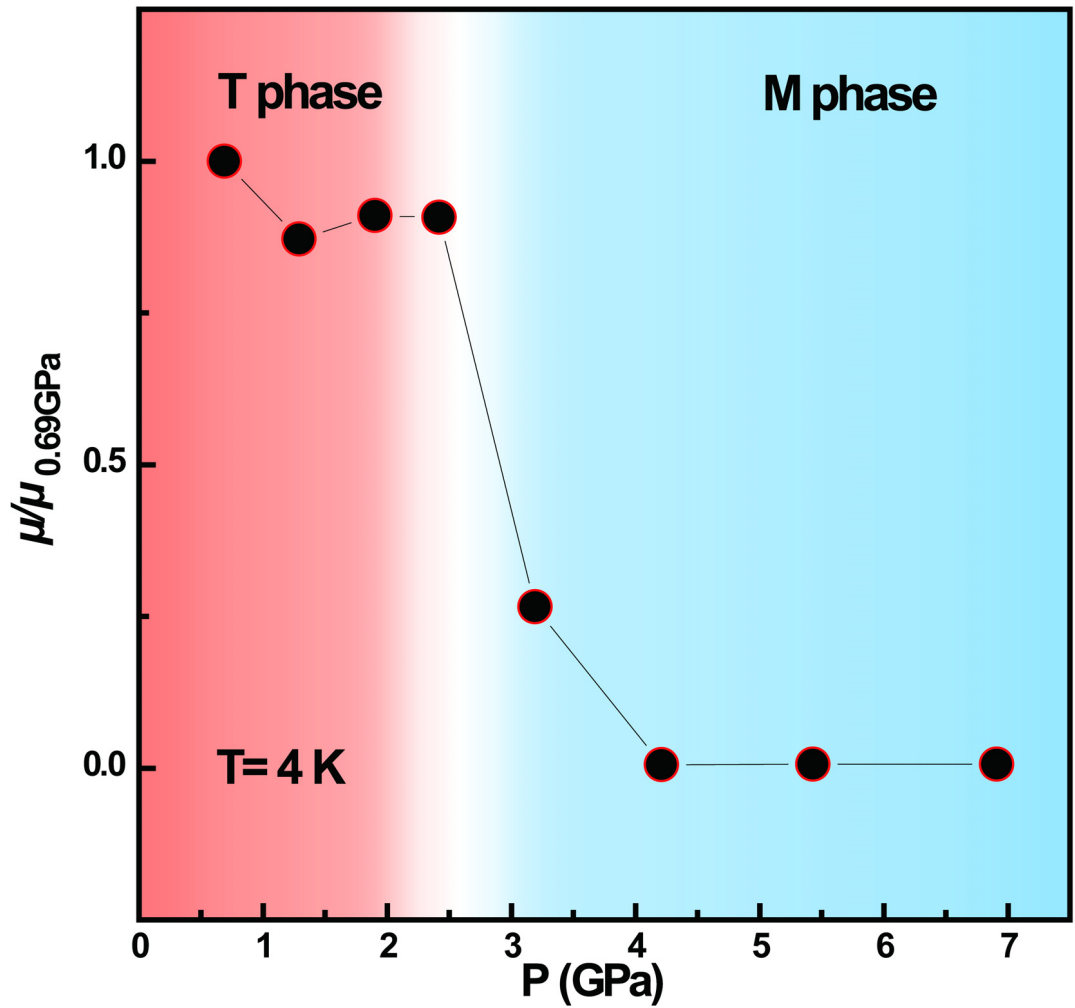

Figure 4 Pressure dependence of the electron mobility of single crystal $Cd_3As_2$ at different temperatures. The data are derived from Hall measurements. The magnetic field is applied perpendicular to the (112) plane, and the current is applied in the (112) plane. The ambient-pressure electron mobility ($\mu$) of the $Cd_3As_2$ sample employed is about $9.2 \times 10^4$ $cm^2/Vs$, in good agreement with values reported [12-18]. The normalized $\mu$ of pressurized $Cd_3As_2$ is derived from the equation of $\mu = R_H/\rho$, where $R_H$ is Hall coefficient and $\rho$ is resistivity.